\documentclass[aps,jcp,superscriptaddress,amsmath,amssymb]{revtex4-2}

\usepackage[utf8]{inputenc}
\usepackage{amsmath}
\usepackage{comment}
\usepackage{mathtools}
\usepackage{xcolor}
\usepackage[T1]{fontenc}
\usepackage[unicode]{hyperref}
\hypersetup{
   unicode=true,          
   plainpages=false,
   colorlinks=true,       
   linkcolor=blue,          
   citecolor=blue,        
}
\usepackage{url}

\newcommand{\bos}[1]{\boldsymbol{#1}}

\def\Eh{E_\text{h}}

\def\br{\boldsymbol{r}}
\def\iim{\text{i}}
\def\tT{\text{T}}

\def\epsi{\varepsilon}

\def\nel{n_\text{el}}

\def\ael{\text{a}}
\def\bel{\text{b}}
\def\cel{\text{c}}
\def\Bbel{\text{B}}
\def\Ccel{\text{C}}

\def\atSup{$\text{a}\ ^3\Sigma_\text{u}^+$}

\def\btPg{$\text{b}\ ^3\Pi_\text{g}$}
\def\ctSgp{$\text{c}\ ^3\Sigma_\text{g}^+$}
\def\BsPg{$\text{B}\ ^1\Pi_\text{g}$}
\def\CsSgp{$\text{C}\ ^1\Sigma_\text{g}^+$}

\def\Eh{E_\text{h}}

\def\br{\boldsymbol{r}}
\def\iim{\text{i}}

\def\nnuc{N_\text{nuc}}
\def\tms{t}

\def\MV{\text{MV}}
\def\Done{\text{D1}}
\def\Dtwo{\text{D2}}
\def\OO{\text{OO}}

\def\bs{\boldsymbol{s}} 

\def\hbp{\hat{\boldsymbol{p}}}
\def\hbs{\hat{\boldsymbol{s}}} 

\def\br{\boldsymbol{r}}

\def\bRnuc{\boldsymbol{R}}

\def\bA{\boldsymbol{A}}
\def\hH{\hat{H}}
\def\nperm{N_\text{perm}}

\def\som{Supporting Information}

\definecolor{ao}{rgb}{0.0, 0.5, 0.0}

\usepackage{setspace}

\DeclareUnicodeCharacter{2212}{-}

\usepackage[unicode]{hyperref}
\usepackage{soul}
\hypersetup{
   unicode=true,          
   plainpages=false,
   colorlinks=true,       
   linkcolor=black,          
   linkcolor=blue,          
   citecolor=blue,        
   urlcolor=blue           
}

\begin{document}

\title{%
Spin-dependent terms of the Breit-Pauli Hamiltonian evaluated with an explicitly correlated Gaussian basis set for molecular computations
}
\author{Péter Jeszenszki}
\affiliation{MTA–ELTE `Momentum' Molecular Quantum electro-Dynamics Research Group,
Institute of Chemistry, Eötvös Loránd University, Pázmány Péter sétány 1/A, Budapest, H-1117, Hungary}

\author{Péter Hollósy}
\affiliation{MTA–ELTE `Momentum' Molecular Quantum electro-Dynamics Research Group,
Institute of Chemistry, Eötvös Loránd University, Pázmány Péter sétány 1/A, Budapest, H-1117, Hungary}

\author{Ádám Margócsy}
\affiliation{MTA–ELTE `Momentum' Molecular Quantum electro-Dynamics Research Group,
Institute of Chemistry, Eötvös Loránd University, Pázmány Péter sétány 1/A, Budapest, H-1117, Hungary}

\author{Edit Mátyus}
\email{edit.matyus@ttk.elte.hu}
\affiliation{MTA–ELTE `Momentum' Molecular Quantum electro-Dynamics Research Group,
Institute of Chemistry, Eötvös Loránd University, Pázmány Péter sétány 1/A, Budapest, H-1117, Hungary}

\date{27 August 2025}

\begin{abstract}
\noindent
This work collects the spin-dependent leading-order relativistic and quantum-electrodynamical corrections for the electronic structure of atoms and molecules within the non-relativistic quantum electrodynamics framework.
We report the computation of perturbative corrections using an explicitly correlated Gaussian basis set, which allows high-precision computations for few-electron systems.
In addition to numerical tests for triplet Be, triplet $\mathrm{H}_2$, and triplet $\mathrm{H}_3^+$ states and comparison with no-pair Dirac--Coulomb--Breit Hamiltonian energies, numerical results are reported for electronically excited states of the helium dimer, He$_2$, for which the present implementation delivers high-precision magnetic coupling curves necessary for a quantitative understanding of the fine structure of its high-resolution rovibronic spectrum.
\end{abstract}

\maketitle

\section{Introduction}
Few-electron atoms and molecules are sometimes referred to as `calculable' systems, which allow to extensively test and extend the frontier of current methodologies and probe small physical effects by comparison with high-resolution spectroscopy experiments.
At the extreme of precision physics applications, the most accurate experimental and most possible complete theoretical treatments for ionization, dissociation and rovibrational energies progress head-to-head \cite{ShBaReHoCa23,ClScAgScMe23,AlGiCoKoSc20}, and ultimately deliver more precise values for fundamental physical constants and set bounds on possible new types of forces~\cite{PaGeKaHaHiKoCoEiUbKo20,GePaKaHiKoSaEiUbKo21}. 
Furthermore, the reliable identification of quantum states is crucial for the manipulation of atoms and molecules with laser light, which contributes to advancements in the field of ultracold molecules \cite{LaVaWaYe24} and quantum technologies \cite{FuWuIsWaMaNaSc24,RuHeGuCo25}.
Theoretical and computational progress in precision physics delivers benchmark values 
for quantum chemistry methods targeting larger systems.

For few-particle systems, an explicitly correlated basis set is frequently used in combination with a variational procedure to capture the essential electron-electron correlation effects, leading to fast convergence in the energy and other physical properties \cite{SuVaBook98,MiBuHoSuAdCeSzKoBlVa13}. To obtain analytic expressions for the matrix elements, floating Explicitly Correlated Gaussians (fECGs) are used as a spatial basis set \cite{SuVaBook98,MiBuHoSuAdCeSzKoBlVa13,St14,St19,Ce03}, for which numerical efficiency in the high-precision computation of few-electron systems has already been demonstrated \cite{PaAd12,TuPaAd12,FeMa19HH,FeKoMa20,PaYePa21,NaTuStKeAd24,FeJeMa22b,MaFeJeMa23,SaFeMa23,FeMa22bethe,FeMa22h3,paper-he2p,paper1,paper2}. 

Since the electronic energy of few-particle systems can be converged to high precision in a variational fECG approach, the relativistic and quantum electrodynamical (QED) contributions become `visible' and important even for (compounds of) light elements. High-resolution spectroscopic measurements reveal small magnetic effects, seen as fine and hyperfine splittings in the spectra. We will refer to the relevant interactions as `spin-dependent' relativistic (and QED) interactions, and we will also use the term `spin-independent' relativistic and QED corrections, which contribute to the centroid energy (defined as the average of the fine-structure energy levels corresponding to the degenerate non-relativistic subspace). 
While the computation of the centroid corrections is well elaborated, including the regularisation techniques \cite{PaCeKo05,JeIrFeMa22,RaFeMaMa24}, which enhance the convergence of singular terms in the Gaussian basis representation (for the incorrect electron-nucleus and electron-electron coalescence behaviour). 

In this paper, we focus on an fECG-based implementation of the spin-dependent matrix elements of the Breit-Pauli Hamiltonian. We also include corrections for the anomalous magnetic moment of the electron, which constitutes the leading-order ($\alpha^3\Eh$, with the $\alpha$ fine-structure constant) QED corrections to the spin-dependent terms. 
We note that higher-order QED ($\alpha^4\Eh$) corrections have been derived by Douglas and Kroll in 1974 for the helium atom triplet states \cite{DoKr74} and more recently, the $\alpha^5\Eh$-order corrections have been derived and evaluated for triplet helium states \cite{PaYePa21,PaYePa24}. Inner shell transitions are in excellent agreement with experiment, though there is a significant deviation for the ionization energy between theory \cite{YePaPuPa20,YePaPa21} and experiment \cite{BeBavaLuMaMcOBRoSaWeWeChEy98,KaGoPiUbEi11,ClJaScAgScMe21,ClMe25}. 

The present work focuses on the relativistic and leading-order QED corrections, \emph{i.e.,} up to $\alpha^3\Eh$ order, but goes beyond atomic applications; molecular systems with clamped nuclei require the development of a floating ECG methodology, reported in the present work.

Non-floating, spherically symmetric, ECG basis sets have already been successfully used in spin-dependent atomic computations, considering beryllium, boron and carbon atomic states   \cite{PuKoPa21,KeStAd20,StKeAd22,YePaPa22,NaTuStKeAd24,NaBuAd24}.  By relaxing the Born–Oppenheimer approximation, small molecules (H$_2^+$ \cite{KoHiKa06,HaKoHiKaH2p22}, HD$^+$ \cite{HaKoHiKaHDp22}, BH$^+$ and BH \cite{NaBuAd25}) have also been computed (with varying precision). However, these approaches cannot be applied directly to clamped nuclei and molecular computations exploiting the Born--Oppenheimer approximation, which would otherwise be advantageous, since the computational complexity grows rapidly with the number of particles.

In this paper, we report a general $N$-electron implementation without restrictions on the number and spatial arrangement of the clamped atomic nuclei. 
The implementation is based on a spinor basis representation, in which the spatial and spin components are evaluated independently and subsequently combined via calculating simple tensor products. The approach is, in principle, `general' and not restricted by the number of electrons or point-group symmetry. For testing the developed methodology and computer implementation, we perform computations for triplet $\mathrm{Be}$ for comparison with literature values. As to molecular systems (the main target of this work), two-electron triplet $\mathrm{H}_2$ and also two-electron triplet $\mathrm{H}_3^+$ (both a linear and a triangular configuration) are also included in the test set. For molecular systems, high-accuracy benchmark values are not available in the literature; so, we tested these results with our in-house developed no-pair Dirac–Coulomb–Breit methodology (currently available for two-spin-1/2-fermion systems) \cite{JeMa23}, in which the spin-dependent Breit-Pauli relativistic effects appear at lowest order of the fine-structure constant.
Then, as a first large-scale application of the implementation, motivated by recent and ongoing experiments,\cite{HoWiShBeMe25,WiHoMe25,VeZeKnRoBe25} we compute magnetic coupling curves for several electronically excited states of the (triplet and singlet) He$_2$ molecule, which significantly improve upon previously available quantum chemistry computations \cite{BeNiMu74,ChJeYaLe89,Ya89,BjMiPaRo98,Mi03,NiKrPrViWi2019,XuLuZhGuSh24}. 
The newly-computed coupling curves are used in rovibronic-fine-structure computations reported in separate papers~\cite{paper1,paper2}.

%
%
\section{Theoretical framework}
For atoms and molecules with low nuclear charge, the non-relativistic description often provides a good starting point. The non-relativistic electronic energy is incremented with relativistic and QED corrections arranged according to powers of the $\alpha$ fine-structure constant. 
In this paper, we include all terms up to $\alpha^3$ (in Hartree atomic units),
\begin{align}
    E=E^{(0)} + \alpha^2 E^{(2)} + \alpha^3 E^{(3)} \ .
\end{align}
We start by solving the Schrödinger equation, 
\begin{align}
   \hH^{(0)} | \varphi^{(0)} \rangle 
   = 
   E^{(0)} | \varphi^{(0)} \rangle 
\end{align}
for the non-relativistic electronic Hamiltonian, 
\begin{align}
  \hat{H}^{(0)}
  &= 
  -\frac{1}{2} \sum_{i=1}^{\nel} \bos{\Delta}_{\br_i} 
  -\sum_{i=1}^{\nel}\sum_{A=1}^{\nnuc}\frac{Z_A}{r_{iA}}
  + 
  \sum_{i=1}^{\nel}\sum_{j>i}^{\nel} \frac{1}{r_{ij}}
  \; , 
\end{align}
with the nuclei clamped at the $\bos{R}_A\ (A=1,\ldots,N_\text{nuc})$ positions. We use the short notation, 
$r_{iX}=|\br_{iX}|$, where $X$ is either a nucleus or an electron index, \emph{i.e.,} resulting in $\br_{iA}=\br_{i}-\bRnuc_{A}$ or $\br_{ij}=\br_{i}-\br_{j}$.

The leading-order relativistic corrections, $\alpha^2 E^{(2)}$, can be computed as the expectation value of the Breit-Pauli (BP) Hamiltonian with the non-relativistic wave function, $\varphi^{(0)}$. 
The focus of this work is the computation of the spin-dependent contributions, so we write the Hamiltonian as the sum of spin-independent (sn) and spin-dependent (sd) terms \cite{DyFaBook07,ReWoBook15}
 \begin{align}
  \hH^{(2)}
  =
  \hH^{(2)}_\text{sn} + \hH^{(2)}_\text{sd} 
  =
  \hH_\MV + \hH_\Done + \hH_\OO + \hH_\Dtwo 
  + \hH_\text{SO}
  + \hH_\text{SOO} + \hH_\text{SOO'}
  + \hH_\text{SS}   \ ,
\end{align}
where the manifestly spin-independent terms are re-iterated only for completeness, \emph{i.e.,}
the mass-velocity (MV), the one-electorn Darwin (D1), the two-electron Darwin (D2), and the orbit-orbit terms (OO) are
 \begin{align}
  \hH_\MV
  =
  -\frac{1}{8} 
  \sum_{i=1}^{\nel} (\hbp_{i}^{2})^2 \; ,
  \quad
  \hH_\Done
  =
  \frac{\pi}{2} \sum_{i=1}^{\nel}
  \sum_{A=1}^{\nnuc}
    Z_A\delta(\br_{iA}) \; ,
  \quad
  \hH_{\Dtwo}
  =
  -\pi
  \sum_{i=1}^{\nel}\sum_{j>i}^{\nel}
    \delta(\br_{ij}) \ ,
\end{align}
and
\begin{align}
  \hH_{\OO}
  =
  -\frac{1}{2}
  \sum_{i=1}^{\nel}\sum_{j>i}^{\nel}
    \left[%
      \frac{1}{r_{ij}}\hbp_{i}\hbp_{j} + \frac{1}{r_{ij}^{3}}(\br_{ij}(\br_{ij}\hbp_{j})\hbp_{i})
    \right] \ ,
\end{align}
respectively.
The corrections carrying spin operators in their expressions are 
the spin-orbit interaction, with $\hat{\boldsymbol{l}}_{iX}=\boldsymbol{r}_{iX}\times \hat{\bos{p}}_i$ and $\hat{\boldsymbol{s}}_i=\frac{1}{2} I(1)\otimes ...\otimes \boldsymbol{\sigma}(i) \otimes...\otimes I(n_\text{el})$ with the $\bos{\sigma}=(\sigma_x,\sigma_y,\sigma_z)$ Pauli matrices, 
\begin{align}
  \hat{H}_{\text{SO}}
  &=
  \frac{1}{2}
  \sum_{i=1}^{\nel} 
  \sum_{A=1}^{\nnuc}
    \frac{Z_A}{r_{iA}^3}
    \hat{\boldsymbol{s}}_i \hat{\boldsymbol{l}}_{iA} \ , \label{eq:defso}
\end{align}
the spin-own-orbit interaction,
\begin{align}
  \hat{H}_{\text{SOO}}
  &=
  -\frac{1}{2}
  \sum_{i=1}^{\nel}\sum_{j>i}^{\nel}
    \frac{1}{r_{ij}^3}
    \left[
      \hat{\boldsymbol{s}}_i \hat{\boldsymbol{l}}_{ij}
      +
      \hat{\boldsymbol{s}}_j \hat{\boldsymbol{l}}_{ji}
    \right] \ , \label{eq:defsoo1}
\end{align}
the spin-other-orbit interaction,
\begin{align}
  \hat{H}_{\text{SOO}'}
  &=
  -\sum_{i=1}^{\nel}\sum_{j>i}^{\nel}
    \frac{1}{r_{ij}^3} 
    \left[%
      \hat{\boldsymbol{s}}_i \hat{\boldsymbol{l}}_{ji}
      +
      \hat{\boldsymbol{s}}_j \hat{\boldsymbol{l}}_{ij}
    \right] \ , \label{eq:defsoo2}
\end{align}
and the spin-spin interaction, 
\begin{align}
  \hH_\text{SS} 
  &= 
  \sum_{i=1}^{\nel} \sum_{j>i}^{\nel} 
  \left[%
     (\hbs_{i}\hbs_{j}) (\hbp_{i}\hbp_{j} \frac{1}{r_{ij}})
    -(\hbs_{i}\hbp_{j}) (\hbs_{j}\hbp_{i} \frac{1}{r_{ij}})
  \right]   
  =
  \hat{H}_\text{SS,dp} + \hat{H}_\text{SS,c}\ , 
  \label{eq:HSS}
\end{align}
which is written as the sum of the magnetic spin dipole-dipole interaction,
\begin{align}
  \hat{H}_\mathrm{SS,dp}
  =
  \sum_{i=1}^{\nel} \sum_{j>i}^{\nel} 
    \left[%
      \frac{\hbs_{i} \hbs _{j}}{r_{ij}^3}
      -\frac{3\left(\hbs_{i}\bos{r}_{ij}\right) \left(\hbs_{j}\bos{r}_{ij}\right)}{r_{ij}^5}
  \right] \label{eq:defssdp}
\end{align}
and the Fermi contact term for each pair of electrons,
\begin{align}
  \hat{H}_\text{SS,c} 
  =  
  -\frac{8\pi}{3} 
  \sum_{i=1}^{\nel}\sum_{j>i}^{\nel}
    \hbs_i \hbs_j \delta(\br_{ij}) 
  =
  -2 \pi \sum_{i=1}^{\nel} \sum_{j>i}^{\nel} \delta(\br_{ij})    
    \ , \label{eq:defHssc}
\end{align}
where in the last step we used the form of the Fermi contact term over the anti-symmetrized Hilbert subspace.

The spin-independent terms, giving rise to the relativistic correction of the `centroid' are collected as
\begin{align}
  \alpha^2\hat{H}^{(2)}_\text{sn}
  =
  \alpha^2[\hH_\MV + \hH_\Done + \hH_\OO + \hH_\Dtwo + \hat{H}_\text{SS,c}] \; .
\end{align}
The precise computation of expectation values (with Gaussian basis sets) for this part has been discussed in detail elsewhere \cite{PaCeKo05,JeIrFeMa22,RaFeMaMa24}.

The focus of the present work is the implementation of the spin-dependent terms
\begin{align}
  \alpha^2 \hat{H}_\text{sd}^{(2)} 
  &=  
  \alpha^2\left[%
    \hat{H}_\text{SO}
    +
    \hat{H}_\text{SOO}
    +
    \hat{H}_\text{SOO'}
    +
    \hat{H}_\text{SS,dp} 
  \right] \; .
  \label{eq:ds2}  
\end{align}
In addition, we consider the effect of the anomalous magnetic moment of the electron  \cite{Sc48,sucherPhD1958,JeAdBook22} on the spin-dependent terms,  which gives rise to the leading-order QED correction for the spin-dependent part (a brief summary with references is in the SM of Ref.~\citenum{paper1}),
\begin{align}
  \alpha^3 \hat{H}_\text{sd}^{(3)} 
  &=   
  \alpha^2
  \left[%
    \frac{\alpha}{\pi} \hat{H}_\text{SO}
    +
    \frac{\alpha}{\pi} \hat{H}_\text{SOO}
    +
    \frac{1}{2}\frac{\alpha}{\pi}\hat{H}_\text{SOO'}
    +
    \frac{\alpha}{\pi}\hat{H}_\text{SS,dp} 
  \right] \; .
  \label{eq:ds3}
\end{align}
So, up to $\alpha^3\Eh$ order, the spin-dependent terms are
\begin{equation}
 \alpha^2\hat{H}_\text{sd}
 =
 \alpha^2\left[\left(1+\frac{\alpha}{\pi}\right)\hat{H}_\text{SO}
 +
 \left(1+\frac{\alpha}{\pi}\right)\hat{H}_\text{SOO}
 +
 \left(1+\frac{\alpha}{2\pi}\right)\hat{H}_\text{SOO'}
 +
 \left(1+\frac{\alpha}{\pi}\right)\hat{H}_\text{SS,dp}\right] \ . \label{eq:Hsdtotal}
\end{equation}

\section{Implementation}
For the numerical implementation, the non-relativistic wave function is expanded in an explicitly correlated basis set,
\begin{align}
  |\varphi^{(\Gamma,SM_S)}\rangle =  \sum_{\mu=1}^N C_\mu |\phi^{(\Gamma,SM_S)}_\mu \rangle  \ , 
  \label{eq:elstate}
\end{align}
where the coefficients $C_\mu$ are obtained by diagonalising the non-relativistic Hamiltonian matrix. 
We define the basis function as
\begin{align}
  \phi^{(\Gamma,SM_S)}_\mu
  =
  \phi^{(\Gamma,SM_S)}(\br;\bA_\mu,\bs_\mu,\bos{\theta}_\mu)
  =
 \hat{\mathcal{A}}
  \lbrace %
    f^{(\Gamma)}(\br;\bA_\mu,\bs_\mu)
    \chi^{(SM_S)}(\bos{\theta}_\mu)
  \rbrace \ .
\end{align}
$\chi^{(SM_S)}(\bos{\theta}_\mu)$ is a spin function, 
$f^{(\Gamma)}(\br;\bA_\mu,\bs_\mu)$ is a spatial function.  In this work,symmetrized floating explicitly correlated Gaussian functions (fECG) \cite{SuVaBook98,MiBuHoSuAdCeSzKoBlVa13,St19,St14,Ce03} are used as spatial functions, 
\begin{align}
  f_\mu^{(\Gamma)}
  =
  f^{(\Gamma)}(\br;\bA_\mu,\bs_\mu)
  =
    \hat{\mathcal{P}}_\Gamma \exp\left[%
    -(\br-\bs_\mu)^\tT (\bA_\mu\otimes \bos{I}_3) (\br-\bs_\mu)
  \right] \ ,
\end{align}
where $\hat{\mathcal{P}}_{\Gamma}$ is the projector onto the $\Gamma$ irreducible representation (irrep) of the point-group defined by the fixed nuclear skeleton (technical details are reported in some detail in the Supplementary Material of Ref. \citenum{JeMa23}), 
$\bos{I}_3$ is the three-dimensional identity matrix, $\bA_\mu$ and $\bs_\mu$ are non-linear parameters, which are generated and optimised by minimisation of the non-relativistic energy \cite{SuVaBook98,MaRe12,Ma19review,MiBuHoSuAdCeSzKoBlVa13,paper1}. 

$\hat{\mathcal{A}}$ is the anti-symmetrization operator for the electrons, 
\begin{align}
  \hat{\mathcal{A}}
  = 
  (\nperm)^{-\frac{1}{2}}
  \sum_{r=1}^{\nperm} (-1)^{p_r} \hat{P}_r \hat{Q}_r 
  \ , \label{eq:antisym}
\end{align}
where the sum is over all permutations ($\nperm=\nel!$), $p_r$ is the parity (odd or even) of the permutation $r$, and $\hat{P}_r$ and 
$\hat{Q}_r$ are the permutation operators acting on the spatial ($\br$) and the spin ($\bos{\sigma}$) degrees of freedom, respectively. 
We note the quasi-idempotency of the antisymmetrizer, $\hat{\mathcal{A}}^2=(\nperm)^{\frac{1}{2}}\hat{\mathcal{A}}$.

Regarding the spin function $\chi^{(SM_S)}_\mu=\chi^{(SM_S)}(\bos{\theta}_\mu)$, $S$ is the total electron spin quantum number and $M_S$ is the quantum number for the spin projection. 
$S,M_S$ for $\nel$ electrons define a spin subspace, for which the spin functions can be expressed as a linear combination of the elementary spin functions; we call this representation the $\theta$-parameterization\cite{SuVaBook98,MaRe12} (see also \som). In the non-relativistic energy minimization, we also include the $\bos{\theta}_\mu$ parameter vector.

Next, let us consider a general, permutationally invariant operator written as a linear combination (in terms of electronic and Cartesian indexes) of products of $\hat{G}_{ia}$ spatial and $\hat{T}_{ia}$ spin-dependent factors,
\begin{align}
  \hat{\mathcal{H}} = \sum_{i=1}^{\nel}\sum_{a=1}^3 \hat{G}_{ia} \hat{T}_{ia} \; .
\end{align}
We compute matrix elements of this operator with non-relativistic electronic functions, $\varphi=\varphi^{(\Gamma,SM_S)}$ and $\varphi'=\varphi^{(\Gamma',S'M_S')}$ as
\begin{align}
  \langle %
    \varphi | \hat{\mathcal{H}} | \varphi'
  \rangle
  &=
  \sum_{\mu=1}^{N} \sum_{\nu=1}^{N'}
    C^\ast_\mu C'_\nu
    \langle 
      \phi_\mu | \hat{\mathcal{H}} | \phi'_\nu
    \rangle
  \nonumber \\
  &=
  \sum_{\mu=1}^{N} \sum_{\nu=1}^{N'}
    C^\ast_\mu C'_\nu
    \langle 
      \hat{\mathcal{A}}\lbrace f_\mu\chi_\mu \rbrace 
      | \hat{\mathcal{H}} |
      \hat{\mathcal{A}}\lbrace f'_\nu\chi'_\nu \rbrace 
    \rangle  
  \nonumber \\
  &=
  \sum_{\mu=1}^{N} \sum_{\nu=1}^{N'}
    C^\ast_\mu C'_\nu
    \sum_{r=1}^{\nperm}
      (-1)^{p_r}
      \langle %
         f_\mu\chi_\mu 
        | \hat{\mathcal{H}} |
        (\hat{P}_r f'_\nu) (\hat{Q}_r \chi'_\nu) 
      \rangle      
  \nonumber \\
  &=
  \sum_{\mu=1}^{N} \sum_{\nu=1}^{N'}
    C^\ast_\mu C'_\nu
    \sum_{r=1}^{\nperm}
      (-1)^{p_r}
      \sum_{i=1}^{\nel}\sum_{a=1}^3
      \langle %
         f_\mu | \hat{G}_{ia} | \hat{P}_r f'_\nu
      \rangle_{\bos{r}}
      \langle %
        \chi_\mu 
        | \hat{T}_{ia} |
        \hat{Q}_r \chi'_\nu
      \rangle_{\bos{\sigma}} \ .
    \label{eq:Gscalc}
\end{align}
The irreducible representation of spatial symmetry can differ for $f_\mu$ and $f_\nu'$, hence, the spatial-symmetry operations are retained in the bra and the ket functions.
For the computation of the spin-orbit (SO), the spin-own-orbit (SOO), and the spin-other-orbit (SOO$'$) terms, Eqs.~\eqref{eq:defso}--\eqref{eq:defsoo2}, the $\hat{G}_{ia}$ factors are 
\begin{align}
  \text{SO}:\quad
  \hat{G}^\text{SO}_{ia}
  &= 
  \frac{1}{2} \sum_{A=1}^{N_\mathrm{nuc}} \frac{Z_A}{r_{i A}^3} \hat{{l}}_{i A,a}  \ , \label{eq:defGso} \\
  \text{SOO}:\quad
  \hat{G}^\text{SOO}_{ia}
  &= 
  -\frac{1}{4} \sum_{j \neq i}^{n_\mathrm{el}} \frac{1}{r_{ij}^3} \hat{{l}}_{ij,a} \ , \label{eq:defGSOO1}\\
  \text{SOO}':\quad
    \hat{G}^{\text{SOO}'}_{ia}
  &= 
  -\frac{1}{2}\sum_{j \neq i}^{n_\mathrm{el}} \frac{1}{r_{ij}^3} \hat{{l}}_{ji,a} \ , \label{eq:defGSOO2} 
\end{align}
and the $\hat{T}_{ia}$ factor is simply the spin matrix,
\begin{align}
    \hat{T}_{ia} =\hat{s}_{ia} \ .
\end{align}
The spin-spin dipolar term, Eq.~\eqref{eq:defssdp}, is a two-electron term, 
and by analogous calculation to Eq.~\eqref{eq:Gscalc},
we obtain
\begin{align}
  \langle %
    \varphi | \hat{H}_\text{SS,dp} | \varphi'
  \rangle
  &=
  \sum_{\mu=1}^{N} \sum_{\nu=1}^{N'}
    C^\ast_\mu C'_\nu
    \sum_{r=1}^{\nperm}
      (-1)^{p_r}
      \sum_{i=1}^{\nel} \sum_{j>i}^{\nel}\sum_{a,b=1}^3
      \langle %
         f_\mu | \hat{G}_{ij,ab} | \hat{P}_r f'_\nu
      \rangle_{\bos{r}}
      \langle %
        \chi_\mu 
        | \hat{s}_{ia}\hat{s}_{jb} |
        \hat{Q}_r \chi'_\nu
      \rangle_{\bos{\sigma}} \; ,
    \label{eq:Gssdpcalc}  
\end{align}
where
\begin{align}
    \hat{G}_{ij,ab} = \frac{\delta_{ab}}{r_{ij}^3} - \frac{3 r_{ij,a}r_{ij,b}}{r_{ij}^5}\ .
\end{align}
The spatial integrals are calculated analytically by exploiting the mathematical properties of fECGs, similarly to our previous work, \emph{e.g.,} Ref.~\citenum{FeKoMa20,JeFeMa21,JeFeMa22,FeJeMa22,FeJeMa22b}.

Since the present work is about the evaluation of the spin-dependent Breit-Pauli matrix elements, some notes regarding the spinor structure and matrix elements are appropriate. 
To evaluate the spin matrix elements, the spin-adapted $\theta$-parametrization (see also Refs.~\citenum{paper-he2p} and \citenum{paper1}) is transformed into the spinor (vector) representation (see also \som),
\begin{align}
  |\chi_\mu^{(SM_S)}(\boldsymbol{\theta})\rangle
  = 
  \sum_{\substack{ \tms_{1},\tms_{2},...,\tms_{{\nel}} \\ \Sigma \tms_{i}=2M_S+1}}
    d_{\tms_{1} \tms_{2} ... \tms_{\nel},\mu}^{(SM_S)} 
    |\Xi_{\tms_{1} \tms_{2} ... \tms_{{\nel} }}\rangle \ , 
    \label{eq:spinor}
\end{align}
where $\tms_i=2 m_{s_i} + 1$, $|\Xi_{\tms_{1} \tms_{2} ... \tms_{\nel}}\rangle=| \eta_{\tms_{1}} \rangle \otimes 
    | \eta_{\tms_{2}} \rangle \otimes \dots \otimes
    | \eta_{\tms_{\nel}} \rangle$,
    and
    $|\eta_{\tms_i}\rangle$ is a single-particle spin function (vector). 
Depending on the $m_{s_{i}}=+\frac{1}{2}$ or $-\frac{1}{2}$ value, the spin function is `spin up',
$|\eta_1\rangle=|\alpha\rangle=\left(\begin{array}{c} 1\\ 0 \end{array}\right)$, 
or `spin down',  
$|\eta_0\rangle=|\beta\rangle=\left(\begin{array}{c} 0\\ 1 \end{array}\right)$. 

Then, the $\hbs_i$ spin operator is represented as the tensor product of (two-by-two) identity matrices and the Pauli matrix,
\begin{align}
 {\bos{s}}_{ia} 
 = 
 \frac{\hbar}{2} \, \bos{I}(1) \otimes \bos{I}(2) \dots \bos{I}(i-1) \otimes \bos{\sigma}_{a}(i) \otimes \bos{I}(i+1) \dots  \otimes \bos{I}(n_\text{el}) \ . \label{eq:defsia}
\end{align}
The matrix representation of the $\hat{s}_{ia} \hat{s}_{jb}$ operator is obtained as $\bos{s}_{ia} \bos{s}_{jb}$ (the spin space is complete), and 
\begin{align}
 \bos{s}_{ia} \bos{s}_{jb} 
 &= 
 \frac{\hbar}{2} \, 
 \bos{I}(1) \otimes 
 \dots 
 \otimes \bos{\sigma}_{a}(i) \otimes 
 \ldots
 \otimes \bos{\sigma}_{b}(j) \otimes  
 \dots  \otimes \bos{I}(n_\text{el}) \;,\quad i\neq j \\
 \bos{s}_{ia} \bos{s}_{ib} 
 &= 
 \frac{\hbar}{2} \, 
 \bos{I}(1) \otimes 
 \dots 
 \otimes \bos{\sigma}_{a}\bos{\sigma}_{b}(i) \otimes 
 \ldots
 \otimes \bos{I}(n_\text{el}) \; .
\end{align}
Then, the matrix elements of the spin functions are written as (where the tensor structure can be further exploited)
\begin{align}
  \langle \chi_\mu | \hat{s}_{ia} | \chi'_\nu \rangle_{\bos{\sigma}}
  &=
  \langle \chi_\mu^{(SM_S)} | \hat{s}_{ia} | \chi_\nu^{(S'M_S')} \rangle_{\bos{\sigma}} 
  = 
  \bos{d}_\mu^{(SM_S)\dagger} \, \bos{s}_{ia} \, \bos{d}_\nu^{(S'M_S')}  \ , \\
  \langle \chi_\mu | \hat{s}_{ia} \hat{s}_{jb} | \chi'_\nu \rangle_{\bos{\sigma}}
  &=
  \langle \chi_\mu^{(SM_S)} | \hat{s}_{ia} \hat{s}_{jb} | \chi_\nu^{(S'M_S')} \rangle_{\bos{\sigma}} 
  = 
  \bos{d}_\mu^{(SM_S)\dagger} \, \bos{s}_{ia} \, \bos{s}_{jb} \, \bos{d}_\nu^{(S'M_S')}  \ .
\end{align}

In a nutshell, the following algorithm has been implemented in our in-house developed fECG-based computer program, named QUANTEN, to compute matrix elements of spin-dependent relativistic operators connecting various electronic-spin states of small molecules (and also atoms, of course):
\begin{enumerate}
  \item %
    The non-relativistic energies and wave functions are determined by the non-linear optimization of the $\bA_\mu$, $\bs_\mu$, and $\bos{\theta}_\mu$ parameters and diagonalization of the Hamiltonian for a specific spin state, \emph{e.g.,} $M_S=0$.
  \item %
    The spin function $\chi_\mu^{(SM_S)}$ with parameters $\bos{\theta}_\mu$ are transformed to the spinor (vector) representation, Eq.~\eqref{eq:spinor}, to obtain the $d_{\tms_{1} \tms_{2} ... \tms_{\nel},\mu}^{(SM_S)}$ coefficients.
  \item %
    The spin operators (matrices) $\bos{s}_i$ are constructed according to Eq.~\eqref{eq:defsia}, and the ladder operators (matrices) are generated as
   \begin{align}
     \bos{s}^{\pm}_i
     &= 
     \bos{s}_{i,x} \pm \iim \bos{s}_{i,y} \ , \quad
     \bos{S}^{\pm}=\sum_{i=1}^{n_\mathrm{el}} \bos{s}^{\pm}_i \ . 
   \end{align}
   Then, the target $M_S$ states are obtained by matrix-vector multiplication,
   \begin{align}
       \bos{d}_\mu^{(S,M_S \pm 1)}=\bos{S}^{\pm} \bos{d}_\mu^{(SM_S)} \ .
   \end{align}

  \item %
    The spatial matrix elements, 
    $\langle f_\mu^{(\Gamma)} | \hat{G}^\mathrm{x}_{ia} | \hat{P}_r f_\nu^{(\Gamma')} \rangle_{\bos{r}}$ (x=SO, SOO, SOO$'$)
    and 
    $\langle f_\mu^{(\Gamma)} | \hat{G}^\mathrm{SS}_{ij,ab} | \hat{P}_r f_\nu^{(\Gamma')} \rangle_{\bos{r}}$ (SS,dp), are computed using the analytic fECG integral expressions.
  \item %
    The spatial and spin contributions are combined according to Eqs.~\eqref{eq:Gscalc} and \eqref{eq:Gssdpcalc}, to obtain the relativistic (and QED) coupling matrix elements of the electron-spin states.
\end{enumerate}

\section{Numerical tests of the implementation}
This section presents numerical examples used to test the computer implementation of the spin-dependent Breit-Pauli Hamiltonian (BP) matrix elements. For the (atomic) test systems, relatively small basis sets are used, which can be readily extended for better-converged results (if required). The value $\alpha^{-1}=137.$035~999~177, recommended by CODATA 2022 \cite{codata22}, is used in all computations.

\subsection{Atomic test system}
For the spherically symmetric (atomic) case, there are high-accuracy ECG implementations already available in the literature, including the analytic angular prefactor corresponding to higher angular momentum states \cite{PuKoPa21,KeStAd20,StKeAd22}. As a first test case of our implementation, intended for the more general molecular problem, we computed the spin-dependent BP matrix elements of the 2 $^3P^\mathrm{o}$ state of the Be atom. 
We computed the expectation value of the spin-dependent BP operator terms, Eq.~\eqref{eq:ds2}, labelled as $\langle \hat{H}_\text{X}\rangle_J\ (\text{X}=\text{SO, SOO \& SOO', SS,dp})$ with the $M_J=0$ wave function (Table~\ref{tab:Be}).
The relative error due to the finite basis size is approximately the same for all three terms; the deviations from the reference values are attributed to the incompleteness of our basis set.
We also note that the expectation value for other $M_J=\pm 1,\pm 2$ values of the $2\ ^3P$ spin subspace of this atomic system can be calculated from the $\langle \hat{H}_\text{Y}\rangle_0$ matrix elements through known relations, \emph{e.g.,} in Ref. \citenum{KeStAd20}.

\begin{table}[h!]
  \centering
  \caption{%
  Be $2\ ^3P^\mathrm{o}$ state: testing the implementation of the spin-dependent Breit--Pauli matrix elements. The non-relativistic energy, $E^{(0)}$ in $E_\mathrm{h}$, and the expectation value of spin-dependent operators for the $M_J=0$ state in $\mu E_\mathrm{h}$, are given as a function of the number of ECG functions, $N$. 
  }
  \begin{tabular}{@{}ll ll ll ll l@{}}  
    \hline\hline\\[-0.475cm]
     $N$ &&
     $E^{(0)}$ && 
     $\alpha^2 \langle \hat{H}_\text{SO}\rangle_0$ && 
     $\alpha^2 \langle \hat{H}_\text{SOO}+ \hat{H}_\text{SOO'}\rangle_0$ &&  
     $\alpha^2 \langle \hat{H}_\mathrm{SS,dp}\rangle_0$ \\     
    \hline
      5  && --14.44          && --24       &&  17.8     && 1.337 \\
     10  && --14.54          && --28       &&  20.6     && 1.363 \\
     20  && --14.553         && --30.4     && 21.76    && 1.439 \\
     30  && --14.561 2       && --31.60    && 22.14 1  && 1.378 \\ 
     50  && --14.565 14      && --32.06    && 22.28 8  && 1.372 \\
     100 && --14.566 69      && --32.18    && 22.27 8  && 1.367 \\
    \hline\\[-0.475cm]
     Ref.~\citenum{StKeAd22} 
        && --14.567 244 230 && --32.24 16 && 22.24 50 && 1.365 \\
    \hline\hline
    \end{tabular} \\
    \label{tab:Be}
\end{table}

\subsection{Molecular applications}
Regarding molecular systems within the Born-Oppenheimer approximation, we are not aware of any literature data for the high-precision computation of the spin-dependent BP matrix elements. 
Still, to be able to test the spin-dependent BP implementation reported in this work, we used our in-house implementation of the high-precision no-pair Dirac–Coulomb–Breit (DCB) approach \cite{JeFeMa21,JeFeMa22,FeJeMa22b,MaFeJeMa23,JeMa23,HoJeMa24}. In particular, Ref.~\citenum{JeMa23} is most relevant to this work. For the positive-energy projection, we employed the `cutting' projector \cite{JeFeMa22} and utilised quadruple precision arithmetic operations. The no-pair DCB approach is currently available only for two-electron systems; however, for two-electron triplet di- and triatomic systems, there is a non-vanishing contribution, which is used as a test case. 
The no-pair DCB energy includes not only the leading-order but also high-order $(Z\alpha)^n$ contributions; but, for small $Z$ nuclear charges, these higher-order contributions are small (so, we did not perform any `$\alpha$' scaling of the variational result \cite{JeFeMa22,FeJeMa22b,NoMaMa24}).

\vspace{0.5cm}
\paragraph{Triplet H$_2$}
The two lowest-energy triplet electronic states of the H$_2$ molecule are labelled as
a~$^3\Sigma_\text{g}^+$ and b~$^3\Sigma_\text{u}^+$. For these states, only the spin-spin matrix elements are non-zero due to spatial symmetry (Table~\ref{tab:H2}). The spin-spin interaction lifts the threefold degeneracy of each state, resulting the $\Delta E^{(2)}= E^{(2)}_0 - E^{(2)}_{\pm 1}$ zero-field splitting, where the subscript of the energy labels the electron spin angular momentum projection on the body-fixed $z$ axis, fixed to the two nuclei. 
There are spin-dependent BP computations in the literature for these states, in which the Full Configuration Interaction (FCI) approach was used and reached 1~m$E_\mathrm{h}$ convergence of the non-relativistic energy \cite{MiLoRiVaAg03}. 
In the present work, we converged the non-relativistic energy with fECGs to 1~$\mu E_\mathrm{h}$. 
The computed $\Delta E^{(2)}$ energy splitting agrees with the literature value (precise to 1-2 digits), but our value is more precise (to 3-4 digits). 
Furthermore, our perturbative BP energy splitting is in excellent agreement with the no-pair DCB splitting. 
The no-pair DCB energy was computed using the same fECG (non-linear) parameterisation as the perturbative computations, and not surprisingly, the two energy splittings converge at a similar pace. 
Differences appear only at the sub-n$E_\mathrm{h}$ ($<10^{-9}\ \Eh$) level, which is well below the finite basis size error (and the range of the leading-order, $\alpha^2\Eh$ relativistic effects).

\begin{table}[h!]
  \centering
  \caption{%
    $\mathrm{H}_2$ a~$^3\Sigma_\text{g}^+$ and b~$^3\Sigma_\text{u}^+$ states ($R=1.4$~bohr): 
    the non-relativistic energy, $E^{(0)}$ in $\Eh$, and the zero-field splitting, $\Delta E$ in  $\mu E_\mathrm{h}$. For the perturbative correction 
    $\Delta E^{\mathrm{BP}}=\alpha^2\Delta E^{(2)} = \alpha^2 [E^{(2)}_{0}-E^{(2)}_{\pm1}]=
    \alpha^2[\langle H_\mathrm{SS,dp} \rangle_0 -  \langle H_\mathrm{SS,dp} \rangle_{\pm 1}]$
    as a function of the number of fECGs, $N$.
    DCB superscript labels the no-pair DCB computations.
  }  
\scalebox{0.95}{%
  \begin{tabular}{@{}ll@{} ll@{} ll@{} ll@{} ll@{} ll@{} l@{}}  
    \hline\hline\\[-0.40cm]
     & 
    \multicolumn{6}{c}{H$_2$ a~$^3\Sigma_\text{g}^+$} && 
    \multicolumn{5}{c}{H$_2$ b~$^3\Sigma_\text{u}^+$} \\
    \cline{3-7} \cline{9-13}\\[-0.40cm]
    $N$ &$\hspace{1cm}$&
    \multicolumn{1}{c}{$E^{(0)}$} && $\Delta E^\mathrm{BP}$ &&
     $\Delta E^\mathrm{DCB}$ &$\hspace{1cm}$& \multicolumn{1}{c}{$E^{(0)}$} && $\Delta E^\mathrm{BP}$ &&  $\Delta E^\mathrm{DCB}$\\     
    \hline\\[-0.40cm]
     10  && --0.713 250    &&  --0.006 1 && --0.006 1 && --0.781 5&& 3.16 && 3.16  \\
     20  && --0.713 553    && \hspace{0.2cm}0.013 9  && \hspace{0.2cm}0.013 9 && --0.783 3 && 3.13 && 3.13  \\
     30  && --0.713 602    && \hspace{0.2cm}0.019 9  && \hspace{0.2cm}0.019 9 && --0.784 121 && 3.040 3 && 3.040 2 \\
     50  && --0.713 635    && \hspace{0.2cm}0.021 45 && \hspace{0.2cm}0.021 45 && --0.784 227 && 3.032 9 && 3.032 9  \\
     100 && --0.713 640 21 && \hspace{0.2cm}0.022 31 && \hspace{0.2cm}0.022 31 && --0.784 243 5 &&  3.031 09 && 3.031 05\\
     150 && --0.713 640 46 && \hspace{0.2cm}0.022 345 && \hspace{0.2cm}0.022 345&& --0.784 244 14&& 3.030 89 && 3.030 84\\
    \hline\\[-0.35cm]
     Ref.~\citenum{MiLoRiVaAg03} 
        && --0.713 0 && \hspace{0.2cm}0.02 && && --0.784 0 && 3.056 & \\
    \hline\hline
    \end{tabular}} \\
    \label{tab:H2}
\end{table}

\vspace{0.5cm}
\paragraph{Triplet H$_3^+$}
For the lowest-energy triplet state of $\mathrm{H}_3^+$, two nuclear geometries are considered (Table~\ref{tab:H3p}). 
Previous FCI computations \cite{SaRoTaAgPa01,CeAlVa03} converged the non-relativistic energy to 10–100~$\mu E_\mathrm{h}$, but we are not aware of any spin-dependent BP computations and zero-field splitting values published in the literature.

We report computations using the spin-dependent BP implementation (presented in this work) and verify the results against our existing no-pair DCB approach at two distinct geometries.
First, we consider the equilibrium structure, which is linear and has two proton-proton distances equal to $R_{\text{pp},1}=R_{\text{pp},2}=2.454$~bohr~\cite{SaRoTaAgPa01}. Due to the triatomic, linear geometry, the threefold degeneracy of the energy is lifted into a non-degenerate and a doubly degenerate pair of states with $M_S=0$ and $M_S=\pm1$ (where we chose the axis defined by the three nuclei for the quantization axis of the electron spin). 
Next, we repeated the computation for another, non-linear geometry (which is near a saddle point~\cite{CeAlVa03} on the potential energy surface) with two proton-proton distances, 
$R_{\text{pp},1}=1.939$ bohr, $R_{\text{pp},2}=5.961$ bohr, and one proton-proton-proton angle 
$\angle_\text{ppp}=64.4^\text{o}$. In this case, the three-fold degeneracy of the energy is completely lifted; so, we report two energy splittings for the three energy levels in Table~\ref{tab:H3p}. 
The spin-dependent BP implementation is in excellent agreement with the results of the no-pair DCB approach. The latter is currently applicable only for two-electron systems, so we proceed with the BP implementation for further, poly-electronic molecular applications (with two or more clamped nuclei).

\begin{table}[h!]
  \centering
  \caption{%
    Triplet $\mathrm{H}_3^+$ ground state: %
    the non-relativistic energy, $E^{(0)}$ in $\Eh$, and the zero-field splitting, $\Delta E$ in $\mu E_\mathrm{h}$, as a function of the number of fECGs, $N$.
    For the linear (equilibrium) structure, the centres of the fECGs were restricted to the nuclear axis and $R_{\text{pp},1}=R_{\text{pp},2}=2.454$~bohr. 
    For the general triangular structure, the centres of the fECGs were restricted in the plane of the triangle (without any further point-group symmetry projections), $R_{\text{pp},1}=1.939$ bohr, $R_{\text{pp},2}=5.961$~bohr, and $\angle_\text{ppp}=64.4^\text{o}$. 
    For the linear (equilibrium) geometry, 
    $\Delta E=E_{0}-E_{\pm1}=\alpha^2[\langle\hat{H}_\text{SS,dp}\rangle_0-\langle\hat{H}_\text{SS,dp}\rangle_{\pm1}]$.  For the triangular structure, $\Delta E_{ij}=E_i-E_j$, where $i$ and $j$ labels the energy levels (please see also text).
    The DCB superscript labels the no-pair Dirac-Coulomb-Breit computation.
    }
\scalebox{0.87}{%
  \begin{tabular}{@{}%
    l@{}l@{\ \ } l@{}l@{\ \ } l@{}l@{\ \ } 
    l@{}l@{\ \ } l@{}l@{\ \ } l@{}l@{\ \ } 
    l@{}l@{\ \ } l@{}l@{\ \ } l@{}}  
    \hline\hline\\[-0.40cm]
     && 
    \multicolumn{5}{c}{Linear (equilibrium) structure} && 
    \multicolumn{9}{c}{Triangular structure} \\
    \cline{3-7} \cline{9-17} \\[-0.40cm]
    $N$ &$\hspace{0.5cm}$&
    \multicolumn{1}{c}{$E^{(0)}$} && $\Delta E^\mathrm{BP}$ &&
     $\Delta E^\mathrm{DCB}$ &$\hspace{1cm}$ & 
     \multicolumn{1}{c}{$E^{(0)}$} && $\Delta E^\mathrm{BP}_{21}$ &&  $\Delta E^{\text{DCB}}_{21}$ && $\Delta E^\mathrm{BP}_{32}$ &&  $\Delta E^\text{DCB}_\mathrm{32}$ \\     
    \hline
      10  && --1.112 7     && 1.781   && 1.781  && --1.097      && 0.005 1   && 0.005 1&& 0.460 2&& 0.460 1 \\
      20  && --1.115 55    && 1.794   && 1.794  && --1.102 6    && 0.006 3   && 0.006 3&& 0.459 56&& 0.459 54 \\
      30  && --1.115 96    && 1.758 0 && 1.758 0 && --1.103 9    && 0.006 9  && 0.006 9&& 0.459 55&& 0.459 53 \\
      50  && --1.116 086   && 1.757 19&& 1.757 23&& --1.104 29   && 0.007 4  && 0.007 4&& 0.459 35&& 0.459 33 \\
      70  && --1.116 102 7 && 1.757 36&& 1.757 29&& --1.104 36   && 0.007 63 && 0.007 63 && 0.459 25 && 0.459 23 \\
     100  && --1.116 107 6 && 1.756 93&& 1.756 86&& --1.104 391  && 0.007 79 && 0.007 79 && 0.459 152&& 0.459 132 \\
     150  && --1.116 108 8 && 1.756 80&& 1.757 12&& --1.104 402 3&& 0.007 871&& 0.007 871 && 0.459 108&& 0.459 086 \\
     200  && --1.116 109 12&& 1.756 72&& 1.756 65&& --1.104 405 5&& 0.007 878&& 0.007 878 && 0.459 097&& 0.459 075 \\
    \hline\\[-0.40cm]
     Ref. 
        && --1.116 106 3 \cite{SaRoTaAgPa01}&& && && --1.104 08 \cite{CeAlVa03}&& && && &&\\
    \hline\hline
    \end{tabular}} \\
    \label{tab:H3p}
\end{table}

\section{Numerical applications: helium dimer}

\begin{figure}
    \centering
    \includegraphics[width=0.55\linewidth]{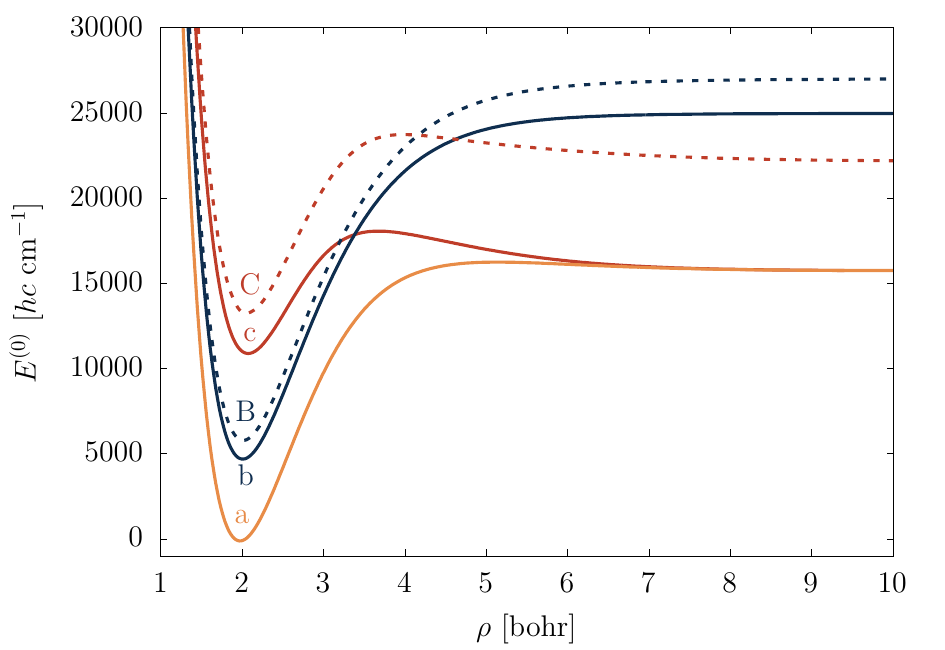}
    \caption{The low-lying excited triplet and singlet states of He$_2$ studied in this work.}
    \label{fig:pec}
\end{figure}

We compute the spin-dependent Breit-Pauli matrix elements connecting electronically excited states of the triplet He$_2$ molecule (Fig.~\ref{fig:pec}). The convergence of the non-relativistic energy at $\rho=2$~bohr (internuclear distance) is reiterated in Table~\ref{tab:energy}; the \atSup\ state basis sets were taken from Ref.~\citenum{paper1} and the \btPg\ and \ctSgp\ states from Ref.~\citenum{paper2}. The \BsPg\ and \CsSgp\ states are computed in this work in order to account for the most important relativistic couplings of the a,b,c subspace. The electronic energy was converged by variational (non-linear) optimisation of the fECG basis functions using the stochastic variational method in combination with the Powell non-linear optimizer~\cite{SuVaBook98,MiBuHoSuAdCeSzKoBlVa13,MaRe12,paper1}.

In what follows, the atomic nuclei are along the (body-fixed) $z$ axis. The $z$ projection of the total, orbital plus electron spin, angular momentum, $\hat{L}_z+\hat{S}_z$ is conserved for clamped nuclei, and its quantum number is labelled with $\Omega$ according to the spectroscopic practice \cite{LBFi04book}. Application of these results in rovibrational and rovibronic computations is reported in Ref.~\citenum{paper1} and also in Ref.~\citenum{paper2}.

In the rest of this section, we report the computation of the relativistic (QED) coupling terms and extensively test their convergence with the basis set size. Comparison with earlier computations by Yarkony~\cite{Ya89} is also shown, although the earlier literature results are not well-converged.

\begin{table}
  \caption{%
    He$_2$ ($\rho=2$ bohr): 
    the non-relativistic energy, $E_x^{(0)}$ in $\Eh$, with respect to the number of optimised fECGs, $N_x$, for the $x=$~a, b, c, B, and C electronic states (Fig.~\ref{fig:pec}). The `a' state basis set is from Ref.~\citenum{paper1}, the `b' and `c' states are from Ref.~\citenum{paper2}.
    \label{tab:energy}
  }
\scalebox{0.85}{%
  \begin{tabular}{@{}  ll@{\ \ \ \ }c  ll@{\ \ \ \ }c  ll@{\ \ \ \ }c ll@{\ \ \ \ }c ll  @{}}
  \hline\\[-0.40cm]
  \hline\\[-0.40cm]
     \multicolumn{2}{c}{\atSup} & \hspace{0.2cm}&  
     \multicolumn{2}{c}{\btPg}  & \hspace{0.2cm}&  
     \multicolumn{2}{c}{\ctSgp}  & \hspace{0.2cm}&       
     \multicolumn{2}{c}{\BsPg}  & \hspace{0.2cm}&       
     \multicolumn{2}{c}{\CsSgp}  \\
     \cline{1-2} \cline{4-5} \cline{7-8} \cline{10-11} \cline{13-14} \\[-0.40cm]
     \multicolumn{1}{c}{$N_\mathrm{a}$} & 
     \multicolumn{1}{c}{$E^{(0)}_\mathrm{a}$} &\hspace{0.2cm}&  
     \multicolumn{1}{c}{$N_\mathrm{b}$} & 
     \multicolumn{1}{c}{$E^{(0)}_\mathrm{b}$} &\hspace{0.2cm}&  
     \multicolumn{1}{c}{$N_\mathrm{c}$} & 
     \multicolumn{1}{c}{$E^{(0)}_\mathrm{c}$} &\hspace{0.2cm}&  
     \multicolumn{1}{c}{$N_\mathrm{B}$} & 
     \multicolumn{1}{c}{$E^{(0)}_\mathrm{B}$} &\hspace{0.2cm}&  
     \multicolumn{1}{c}{$N_\mathrm{C}$} & \multicolumn{1}{c}{$E_\mathrm{C}^{(0)}$} \\
    \hline\\[-0.45cm]
      50 & --5.146 7       & &   50 & --5.124 2       & &   50 & --5.092 0   & & 300 & --5.124 27 & &
      300 & --5.089 38 \\
     100 & --5.149 96      & &  100 & --5.127 87      & &  100 & --5.097 6   & & 500 & --5.124 38 & & 500 & --5.089 739 \\
     200 & --5.150 79      & &  200 & --5.128 822     & &  300 & --5.100 00  & &     &  & & 750 & --5.089 804 \\
     500 & --5.151 071     & &  300 & --5.128 823     & &  500 & --5.100 315 & &     &  & &     & \\
    1000 & --5.151 114     & &  500 & --5.129 246     & &  750 & --5.100 426 & &     &  & &     & \\
    1500 & --5.151 122 5   & &  750 & --5.129 307     & & 1000 & --5.100 465 & &     &  & &     & \\
    2000 & --5.151 123 8   & & 1000 & --5.129 330     & & 1500 & --5.100 482 & &     &  & &     & \\
  \hline\\[-0.4cm]
  \hline
  \end{tabular}
}
\end{table} 

\subsection{Fine structure of the \texorpdfstring{\atSup}\ \ state} 

The lowest-energy triplet state of the $\mathrm{He}_2$ molecule is the a~$^3\Sigma_\text{u}^+$ state. 
The relativistic magnetic spin dipole-dipole interaction, Eq.~\eqref{eq:defssdp}, lifts the threefold degeneracy of this state by splitting it into components with $\Omega=0$ and $\Omega=\pm1$ (Fig.~\ref{fig:Ediag}).
This splitting can be characterized by a single parameter, $\kappa$, which is the energy deviation of the $\Omega=0$ state from the centroid. The corresponding shift of the $\Omega=\pm1$ states is $-\frac{\kappa}{2}$, so the weighted average of the energies equals the centroid energy. The parameter $\kappa$ has relativistic and QED contributions,
\begin{align}
  \kappa
  = \alpha^2 \kappa^{(2)} + \alpha^3 \kappa^{(3)}\ .
  \label{eq:kappa}
\end{align}
Since $\kappa^{(2)}$ originates solely from the relativistic magnetic spin dipolar term (SS,dp), Eq.~\eqref{eq:defssdp}, its QED correction is obtained by simple multiplication, Eq.~\eqref{eq:ds3}, 
\begin{align}
  \kappa^{(3)}= \frac{1}{\pi} \kappa^{(2)} \ .
  \label{eq:kappa3}
\end{align}

For the \atSup\ state, we think that the non-relativistic energy is converged to 1–10~$\mu E_\mathrm{h}$ for the largest basis sets at $\rho=2$~bohr internuclear distance (Table~\ref{tab:energy}). 
The convergence of $\kappa^{(2)}$ with respect to the number of basis functions is shown in Table~\ref{tab:numrelcpl}. For the largest basis sets ($N_\mathrm{a}=1000$–$2000$), its value is converged to at least three significant digits. These results agree to one(-two) digit(s) of the Multireference Configuration Interaction (MRCI) computations reported in the literature \cite{BjMiPaRo98}. The somewhat lower accuracy of the MRCI results is most probably related to the finite basis set error, which we can significantly improve by using an explicitly correlated basis set. 

The A~$^1\Sigma_\text{g}^+$ state is closest to the a-state near the equilibrium structure (Fig.~\ref{fig:pec}), but there is no direct (first-order) coupling between a and A states through the relativistic (and QED) operators, Eqs.~\eqref{eq:Hsdtotal}. Well, they can couple through second-order effects (some details are collected in the \som), exploratory computations show that these $\alpha^4\Eh$ order contributions are tiny (on the few pico–Hartree, $\mathrm{p}E_\mathrm{h}$ level), so the A state is not considered further in this work.

\subsection{Fine structure of the \texorpdfstring{\btPg}\ \ and \texorpdfstring{\ctSgp}\ \ states}
\begin{figure}
  \centering
  \includegraphics[width=0.55\linewidth]{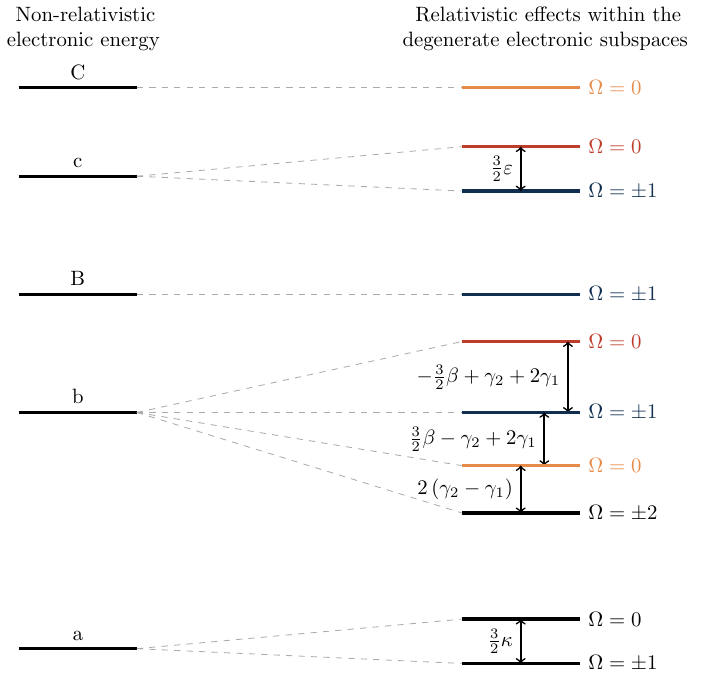}
  \caption{%
    Triplet $\mathrm{He}_2$: splitting of the non-relativistic energies degenerate in $M_S$ by spin-dependent relativistic (and QED) terms. For the definition of $\lambda$, $\delta_1$, $\delta_2$, $\zeta,$ and $\xi$  please see Table~\ref{tab:relcpltotal} (and Ref.~\citenum{paper2}).
    Colour-coding is employed to indicate states that are coupled via relativistic interactions between distinct non-degenerate subspaces (orange: $\lambda$, red: $\delta_2$, blue: $\delta_1, \zeta,\xi$).
    \label{fig:Ediag}    
  }
\end{figure}

The non-vanishing spin-dependent BP matrix elements connecting the b and c states are collected in Table~\ref{tab:relcpltotal}. The parameter $\varepsilon$ labels the separation (zero-field splitting) of the $M_S = 0$ and $M_S = \pm 1$ states in the c subspace. 
For the b subspace, the non-zero electron orbital angular momentum introduces a more complicated splitting pattern. 
It takes three independent parameters, labelled as $\beta$, $\gamma_1$, and $\gamma_2$, which describe the shift and splitting of the originally six-fold degenerate subspace into three two-fold degenerate states with $\Omega = \pm 2$, $ \Omega = \pm 1$, and $ \Omega = 0$ (Fig.~\ref{fig:Ediag}). 

Additionally, the b and c electronic-spin states couple through the matrix elements labelled as $\delta_1$ and $\delta_2$ in Table~\ref{tab:numrelcpl}, hence, only the b states for $\Omega=\pm 2$ remain degenerate. 

Non-negligible couplings also exist with the singlet states, most importantly,  with the \BsPg\ and the \CsSgp\ states (Fig.~\ref{fig:pec}); the non-vanishing BP matrix elements are labelled with $\lambda$, $\zeta$, and $\xi$ in Table~\ref{tab:relcpltotal}.
Similarly to $\kappa$, the relativistic and QED couplings for the parameters $\beta$, $\gamma_1$, and $\epsilon$ differ only by a simple multiplicative factor, since only the magnetic spin dipolar term contributes to the relativistic coupling,
\begin{align}
  \beta^{(3)}
  =
  \frac{1}{\pi} \beta^{(2)}\ , \hspace{2cm} 
  \gamma^{(3)}_{1}
  =
  \frac{1}{\pi}\gamma^{(2)}_{1} \ ,  \hspace{2cm} 
  \epsilon^{(3)}
  =
  \frac{1}{\pi} \epsilon^{(2)} \ . \label{eq:linrelqed}
\end{align}
For the other matrix elements, the QED terms were calculated according to Eq.~\eqref{eq:ds3}, and their convergence is shown in Table~\ref{tab:numrelcplQED}.
For matrix elements with the singlet B and C states, due to the smaller basis sets used in this work, the non-relativistic energies are by 1-2 orders of magnitude less accurate than the a non-relativistic energy (Table~\ref{tab:energy}). Despite this lower accuracy, the precision of the relativistic and QED couplings appears to be comparable to that of $\kappa$, the uncertainties are in the second digit after the decimal point (Tables~\ref{tab:numrelcpl} and \ref{tab:numrelcplQED}). 
To test the convergence, we increased the B and C basis sizes to $N_\text{B} = 500$ and $N_\text{C} = 750$, respectively, while using the largest b and c basis sets. The resulting changes in the coupling values are in the 2nd to the 4th digits. These results are expected to be sufficiently accurate for the rovibronic computations, primarily focusing on b-c states, as presented in Ref.~\citenum{paper2}.

\begin{table}
  \caption{%
    Non-vanishing spin-dependent relativistic (and QED) matrix elements within the b, c, B, C  electronic-spin subspace of He$_2$. The superscript of each state labels the Cartesian component of the spatial part (as computed in QUANTEN) and the $M_S$ quantum number of the total electron-spin projection on the internuclear axis.
    The `$.$' labels zero (0). (Further details can be found in the Supplementary Material of Ref. \citenum{paper2}.) 
    \label{tab:relcpltotal}
  }
\scalebox{0.9}{%
  \begin{tabular}{@{}|l||ccc @{\ }|@{\ } ccc @{\ }|@{\ } ccc | cc |c |@{}}
    \hline\\[-0.45cm]
    & $\bel^{x,-1}$ & $\bel^{x,0}$ & $\bel^{x,1}$ %
    & $\bel^{y,-1}$ & $\bel^{y,0}$ & $\bel^{y,1}$ 
    & $\cel^{0,-1}$ & $\cel^{0,0}$ & $\cel^{0,1}$
    & $\Bbel^{x,0}$ & $\Bbel^{y,0}$ & $\Ccel^{0,0}$ 
    \\
    \hline\hline\\[-0.45cm]
    $\bel^{x,-1}$
    & $-\beta$ & .               & $-\gamma_1$       
    & $-\iim\gamma_2$    & .               & $-\iim\gamma_1$   
    & .                  & $\delta_2$      & .                 
    & .                  & .                                   
    & $\lambda$                                                
    \\ 
    $\bel^{x,0}$
    & .               & $2\beta$         & .                 
    & .               & .               & .                 
    & $-\delta_1$     & .               & $\delta_1$        
    & .               & $\iim \zeta$                        
    & .                                                     
    \\     
    $\bel^{x,1}$
    & $-\gamma_1$     & .               & $-\beta$ 
    & $\iim\gamma_1$  & .               & $\iim\gamma_2$     
    & .               & $-\delta_2$     & .                  
    & .               & .                                    
    & $\lambda$                                              
    \\
    \hline\\[-0.45cm]
    $\bel^{y,-1}$
    & $\iim\gamma_2$     & .               & $-\iim\gamma_1$   
    & $-\beta$ & .               & $\gamma_1$        
    & .                  & $\iim\delta_2$  & .                 
    & .                  &  .                                  
    & $\iim \lambda$                                           
    \\ 
    $\bel^{y,0}$
    & .               & .               & .                 
    & .               & $2\beta$         & .                 
    & $\iim\delta_1$  & .               & $\iim\delta_1$    
    & $-\iim \zeta$   &      .                              
    & .                                                     
    \\     
    $\bel^{y,1}$
    & $\iim\gamma_1$  & .               & $-\iim\gamma_2$    
    & $\gamma_1$      & .               & $-\beta$ 
    & .               & $\iim\delta_2$  & .                  
    & .               & .                                    
    & $-\iim \lambda$                                        
    \\     
    \hline\\[-0.45cm]
    $\cel^{0,-1}$
    & .                  & $-\delta_1$     & .                 
    & .                  & $-\iim\delta_1$ & .                 
    & $-\epsi$ & .               & .                 
    & $\xi$              & $\iim \xi$                         
    & .                                                        
    \\ 
    $\cel^{0,0}$
    & $\delta_2$      & .               & $-\delta_2$       
    & $-\iim\delta_2$ & .               & $-\iim\delta_2$   
    & .               & $2\epsi$         & .                 
    & .               & .                                   
    & .                                                     
    \\ 
    $\cel^{0,1}$ 
    & .               & $\delta_1$      & .                  
    & .               & $-\iim\delta_1$ & .                  
    & .               & .               & $-\epsi$ 
    & $\xi$           & $-\iim \xi$                           
    & .                                                      
    \\ 
    \hline\\[-0.45cm]
    $\Bbel^{x,0}$
    & .               & .               & .                 
    & .               & $\iim\zeta$     & .                 
    & $\xi$           & .               & $\xi$             
    & .        & .                                          
    & .                                                     
    \\         
    $\Bbel^{y,0}$
    & .               & $-\iim\zeta$    & .                 
    & .               & .               & .                 
    & $-\iim\xi$       & .               & $\iim\xi$        
    & .               & .                                   
    & .                                                     
    \\
    \hline\\[-0.45cm]
    $\Ccel^{0,0}$
    & $\lambda $      & .               & $\lambda$         
    & $-\iim\lambda$  & .               & $\iim\lambda$     
    & .               & .               & .                 
    & .               & .                                   
    &  .                                                     
    \\         
    \hline 
  \end{tabular}
}
\end{table}

\begin{table}
  \caption{%
    Convergence of the relativistic spin-dependent contributions, 
    in m$\Eh$, 
    within the $\ael$, $\bel$, $\cel$, B, and C electronic-spin subspace of He$_2$ ($\rho= 2$~bohr). 
    $\kappa$ corresponds to the zero-field splitting of the `a' state, and 
    Table~\ref{tab:relcpltotal} defines all other non-vanishing matrix elements within the b, c, B, C subspace.
    $N_\text{B} = 300$ and $N_\text{C} = 500$, unless stated otherwise.
    \label{tab:numrelcpl}
  }
\scalebox{0.87}{%
  \begin{tabular}{@{} r@{\ }r@{}r@{\ }c|c@{} l@{}c@{} l@{\ }c@{} l@{\ }c@{} l@{\ }c@{}  l@{\ }c@{} l@{\ }c@{} l@{\ }c@{} l@{\ }c@{} l@{\ }c@{} l @{}}
  \hline\hline\\[-0.45cm]
    $N_\mathrm{a}$ & 
    $N_\mathrm{b}$ & 
    $N_\mathrm{c}$ &\hspace{0.2cm}&\hspace{0.2cm}& 
    \multicolumn{1}{c}{$\alpha^2\kappa^{(2)}$} & \hspace{0.5cm} &          
    \multicolumn{1}{c}{$\alpha^2\beta^{(2)}$} & \hspace{0.2cm} & 
    \multicolumn{1}{c}{$\alpha^2\gamma^{(2)}_{1}$} & \hspace{0.2cm} & 
    \multicolumn{1}{c}{$\alpha^2\gamma^{(2)}_{2}$} & \hspace{0.2cm} & 
    \multicolumn{1}{c}{$\alpha^2\delta^{(2)}_{1}$} & &  
    \multicolumn{1}{c}{$\alpha^2\delta^{(2)}_{2}$} & \hspace{0.2cm} & 
    \multicolumn{1}{c}{$\alpha^2\varepsilon^{(2)}$}  & \hspace{0.2cm} & 
    \multicolumn{1}{c}{$\alpha^2\lambda^{(2)}$}  & \hspace{0.2cm} & 
    \multicolumn{1}{c}{$\alpha^2\zeta^{(2)}$}  & \hspace{0.2cm} & 
    \multicolumn{1}{c}{$\alpha^2\xi^{(2)}$}  
    \\
    \hline
    50 & 50 & 50       &&& 1.7   && --3.19   & & 12.24   & & 20.7   & & --1.57 & &  --8.68  & & 1.61  && --5.735 & & 28.831 & & 10.070 \\
    100 & 100 & 100    &&& 1.9   && --3.19   & & 12.43   & & 19.99  & & --1.24 & &  --8.67  & & 1.65  && --5.756 & & 29.606 & & 10.495 \\
    200 & 300 & 300    &&& 2.01  && --3.182  & & 12.472  & & 19.69  & & --1.01 & &  --8.596 & & 1.688 && --5.768 & &  29.815 & & 10.764 \\ 
    500 & 500 & 500    &&& 2.07  && --3.1694 & & 12.4671 & & 19.621 & & --0.931 & & --8.559 & & 1.6947 && --5.768 & & 29.870 & & 10.835 \\ 
    1000 & 750 & 750   &&& 2.078 && --3.1689 & & 12.4691 & & 19.598 & & --0.914 & & --8.549 & & 1.6951 && --5.764 & & 29.881 & & 10.858\\
    1500 & 1000 & 1000 &&& 2.081 && --3.1687 & & 12.4691 & & 19.586 & & --0.901 & & --8.538 & & 1.6946 && --5.763 & & 29.883 && 10.862\\
    2000 & 1000 & 1500 &&& 2.082 && --3.1687 & & 12.4691 & & 19.586 & & --0.898 & & --8.537 & & 1.6952 && --5.763 & & 29.883 && 10.866 \\
    \multicolumn{1}{c}{---} & 1000 & \hspace{0.5cm} 1500$^\dagger$ \hspace{-0.25cm} &&& \multicolumn{1}{c}{---} && \multicolumn{1}{c}{---} & & \multicolumn{1}{c}{---} & & \multicolumn{1}{c}{---} & & \multicolumn{1}{c}{---} & & \multicolumn{1}{c}{---} & & \multicolumn{1}{c}{---} && --5.767 & & 29.929 && 10.874 \\
    \hline 
    \multicolumn{3}{c}{Musher {\it \& co-w.} \cite{BeNiMu74}$^*$} &&& 2.01 && \multicolumn{1}{c}{---} && \multicolumn{1}{c}{---} && \multicolumn{1}{c}{---} && \multicolumn{1}{c}{---} && \multicolumn{1}{c}{---} && \multicolumn{1}{c}{---} && \multicolumn{1}{c}{---} && \multicolumn{1}{c}{---} && \multicolumn{1}{c}{---} \\
    \multicolumn{3}{c}{Minaev \cite{Mi03}$^\circ$} &&& \multicolumn{1}{c}{---} && \multicolumn{1}{c}{---} && \multicolumn{1}{c}{---} && \multicolumn{1}{c}{---} && \multicolumn{1}{c}{---} && \multicolumn{1}{c}{---} && \multicolumn{1}{c}{2.14} && \multicolumn{1}{c}{---} && \multicolumn{1}{c}{---} && \multicolumn{1}{c}{---} \\
    \multicolumn{3}{c}{Yarkony \cite{Ya89}} &&& \multicolumn{1}{c}{---} && \multicolumn{1}{c}{---} && \multicolumn{1}{c}{---} && \multicolumn{1}{c}{---} && --1.20 && --4.39 && \multicolumn{1}{c}{---} && --4.359 && 24.330 && 10.375 \\
     \multicolumn{3}{c}{Rosmus {\it \& co-w.} \cite{BjMiPaRo98}} &&& 2.107 && --3.200 && 12.552 && 20.62 && \multicolumn{1}{c}{---} && \multicolumn{1}{c}{---} && \multicolumn{1}{c}{---} && \multicolumn{1}{c}{---} && \multicolumn{1}{c}{---} && \multicolumn{1}{c}{---} \\
    \hline\hline
  \end{tabular}}\\[0.2cm]
  \begin{flushleft}
  $\dagger:$ $N_\mathrm{B}=500$ and $N_\mathrm{C}=750$.\\
  $*:$ $R=2.015$ bohr bond distance for a~$^3 \Sigma_\text{u}^+$. \\
  $\circ:$ $R=2.08$ bohr bond distance for c~$^3 \Sigma_\text{g}^+$.
  \end{flushleft}
\end{table}

\begin{table}
  \caption{%
    Convergence of the spin-dependent, leading-order QED contributions, in 0.1 m$\Eh$, 
    within the $\bel$, $\cel$, $\Bbel$, and $\Ccel$ electronic-spin subspace of $\mathrm{He}_2$ 
    (see also Tables~\ref{tab:relcpltotal} and \ref{tab:numrelcpl}). 
    $N_\text{B} = 300$ and $N_\text{C} = 500$, unless stated otherwise.
    $\beta^{(3)}$, $\gamma^{(3)}_{1}$, $\varepsilon^{(3)}$, and $\kappa^{(3)}$ can be calculated from Eqs.~\eqref{eq:kappa3}--\eqref{eq:linrelqed} and Table~\ref{tab:numrelcpl}.
    \label{tab:numrelcplQED}
  }
\scalebox{0.95}{%
  \begin{tabular}{@{} rr c|c  lc  lc lc lc lc l @{}}
  \hline\hline\\[-0.45cm]
     $N_\mathrm{b}$ & $N_\mathrm{c}$ &\hspace{0.2cm}&\hspace{0.2cm}&  
     \multicolumn{1}{c}{$\alpha^3 \gamma^{(3)}_{2}$} & \hspace{0.2cm} & 
     \multicolumn{1}{c}{$\alpha^3 \delta^{(3)}_{1}$} & & 
     \multicolumn{1}{c}{$\alpha^3 \delta^{(3)}_{2}$} & \hspace{0.2cm} & 
     \multicolumn{1}{c}{$\alpha^3 \lambda^{(3)}$}  & \hspace{0.2cm} & 
     \multicolumn{1}{c}{$\alpha^3 \zeta^{(3)}$}  & \hspace{0.2cm} & 
     \multicolumn{1}{c}{$\alpha^3 \xi^{(3)}$} \\
  \hline\\[-0.45cm]    
     50 & 50     &&& 0.28  & & --1.22   & & --1.13  & & 1.236 & & --7.66   & & 2.06\\
     100 & 100   &&& 0.02  & & --1.24   & & --0.84  & &  1.250 & & --7.87   & & 2.15\\
     300 & 300   &&& 0.138 & & --1.511  & & --0.91  & & 1.261 & & --7.94   & & 2.221 \\
     500 & 500   &&& 0.160 & & --1.502  & & --0.925 & & 1.264 & & --7.951  & & 2.238\\ 
     750 & 750   &&& 0.168 & & --1.500  & & --0.931 & & 1.2630 & & --7.9551 & & 2.243\\
     1000 & 1000 &&& 0.171 & & --1.4974 & & --0.934 & & 1.2626 & & --7.9560 & & 2.244\\
     1000 & 1500 &&& 0.171 & & --1.4966 & & --0.935 & & 1.2626 & & --7.9560 & & 2.246\\
     1000 & 1500$^\dagger$\hspace{-0.13cm} &&& \multicolumn{1}{c}{---} & & \multicolumn{1}{c}{---} & & \multicolumn{1}{c}{---} & & 1.2638 & & --7.9676 & & 2.2453\\
  \hline\hline
  \end{tabular}
}
  \begin{flushleft}
  \hspace{4cm} $\dagger:$ $N_\mathrm{B}=500$ and $N_\mathrm{C}=750$.
  \end{flushleft}
\end{table}

\section{Summary, conclusion, and outlook}
In this paper, we reported methodological details and benchmark numerical results for the spin-dependent relativistic and leading-order QED couplings of electronic states of small molecules. 
The relativistic and QED couplings of the high-precision non-relativistic electronic states are computed as matrix elements of the spin-dependent operators of the Breit-Pauli Hamiltonian, including the spin-orbit, spin-own-orbit, spin-other-orbit, and spin-spin terms. The corrections due to the electron's anomalous magnetic moment are also accounted for.
To accurately describe small polyelectronic molecules, we use a variational floating explicitly correlated Gaussian (fECG) basis procedure.

For small molecular species, accurate and complete fine-structure data are scarcely available, and this work fills this gap. First of all, we tested the methodology for an atomic system, for which (higher precision) fine-structure splitting data are already available in the literature. 
Then, we computed spin-dependent Breit-Pauli Hamiltonian matrix elements for the two-electron triplet H$_2$ and triplet H$_3^+$. High-precision literature data are not available for these fine-structure splittings; however, for these two-electron systems, we verified the results against our in-house developed no-pair Dirac-Coulomb-Breit (np-DCB) Hamiltonian approach, using the same fECG spatial basis set in the two computations. The two approaches give identical splittings to high precision (the most stringent, direct comparison would be possible by $\alpha$-scaling the np-DCB results and comparing the $\alpha^2\Eh$ order term).

The spin-dependent Breit-Pauli Hamiltonian matrix elements can be evaluated for (in principle) general poly-electronic and poly-atomic molecules, whereas our current np-DCB ECG implementation is for two electrons (but automatically includes higher-order $(Z\alpha)^n$ effects).

As to new numerical results with the developed methodology, we converge the relativistic and QED couplings for electronically excited triplet (\atSup, \btPg, and \ctSgp) and singlet (\BsPg, \CsSgp) states of the four-electron He$_2$ molecule to at least 3-4 significant digits, which significantly improves upon (the scarcely) available data in the literature. These results are computed for a series of nuclear configurations and will be used to compute high-resolution spectra, including modelling the fine structure of the rovibronic transitions in a separate paper.

\vspace{-0.075cm}
\section*{Acknowledgement}
\noindent
PJ gratefully acknowledges the support of the János Bolyai Research Scholarship of the Hungarian Academy of Sciences (BO/285/22). We thank the European Research Council (Grant No. 851421) and the Momentum Programme of the Hungarian Academy of Sciences (LP2024-15/2024) for their financial support. We acknowledge DKP for granting us access to the Komondor HPC facility.

%

\clearpage
\begin{center}
\vspace{0.25cm}
{\large
\textbf{Supporting Information}
}\\[0.75cm]
{\large
\textbf{Spin-dependent terms of the Breit-Pauli Hamiltonian evaluated with an explicitly correlated Gaussian basis set for molecular computations}
} \\[0.5cm]

Péter Jeszenszki,$^1$ Péter Hollósy,$^1$ Ádám Margócsy,$^1$ Edit Mátyus$^{1,\ast}$ \\
\emph{$^1$~MTA–ELTE `Momentum' Molecular Quantum electro-Dynamics Research Group,
Institute of Chemistry, Eötvös Loránd University, Pázmány Péter sétány 1/A, Budapest, H-1117, Hungary} \\
$^\ast$ edit.matyus@ttk.elte.hu 
~\\[0.15cm]
(Dated: 27 August 2025)
\end{center}

\setcounter{section}{0}
\renewcommand{\thesection}{S\arabic{section}}
\setcounter{subsection}{0}
\renewcommand{\thesubsection}{S\arabic{section}.\arabic{subsection}}

\setcounter{equation}{0}
\renewcommand{\theequation}{S\arabic{equation}}

\setcounter{table}{0}
\renewcommand{\thetable}{S\arabic{table}}

\setcounter{figure}{0}
\renewcommand{\thefigure}{S\arabic{figure}}

\section{Representation of the spin function with spin-adapted states and spinors - case study for four electrons with \texorpdfstring{$M_S=0$}\ }\label{sec:SpinFunctions}
The simplest representation of the spin part of the wave function is via spinors, which are 
basis vectors in $\mathbb{C}^{2^N}$,
\begin{align}
  |\Xi_{\tms_1 \tms_2 ...\tms_{\nel}}\rangle
  =| \eta_{\tms_1} \rangle \otimes 
   | \eta_{\tms_2} \rangle \otimes \dots \otimes
   | \eta_{\tms_{\nel}} \rangle \, ,
\end{align}
where $\tms_{i}=2m_{s_i}+1$ and $|\eta_{\tms_{i}}\rangle$ is a single-particle spin function (vector). 
For $\tms_{i}=1$, the spin function is ‘spin up’,
\begin{align}
  |\eta_{1}\rangle
  =
  |\alpha\rangle=\left(\begin{array}{c} 1\\ 0 \end{array}\right) 
  \ ,
\end{align} 
for $\tms_{i}=0$, it is `spin down',  
\begin{align}
  |\eta_{0}\rangle=|\beta\rangle=\left(\begin{array}{c} 0\\ 1 \end{array}\right) 
  \ .
\end{align}
The $\{|\Xi_{\tms_{1} \tms_{2} ...\tms_{\nel}}; t_i=0,1\rangle\}$ set forms a basis set for the spin eigenfunctions (see Ref. \citenum{Pa79book} and the Supplementary Material of Ref.~\citenum{paper1}).
For four electrons with $M_S=0$, the spin-adapted states are as follows (see also Ref.~\citenum{SuVaBook98}). We have two singlet ($S=0$) states,
         \begin{align}
          |\Theta_{0,0}^{(1)}\rangle&=\frac{1}{\sqrt{12}}\big(  |\Xi_{0011}\rangle -2
          |\Xi_{0101}\rangle +
          |\Xi_{0110}\rangle + 
          |\Xi_{1001}\rangle - 
          2|\Xi_{1010}\rangle +
          |\Xi_{1100}\rangle \big) \\
          |\Theta_{0,0}^{(2)}\rangle&=\frac{1}{2}\big( |\Xi_{0011}\rangle - |\Xi_{0110}\rangle -           |\Xi_{1001}\rangle + |\Xi_{1100}\rangle  \big)  \ ,
         \end{align}
three triplet ($S=1$) states,
         \begin{align}
          |\Theta_{1,0}^{(1)}\rangle&=\frac{1}{\sqrt{2}}\big( |\Xi_{1100}\rangle - |\Xi_{0011}\rangle  \big) \\ 
          |\Theta_{1,0}^{(2)}\rangle&=\frac{1}{\sqrt{2}}\big( |\Xi_{1010}\rangle - |\Xi_{0101}\rangle  \big) \\
          |\Theta_{1,0}^{(3)}\rangle&=\frac{1}{\sqrt{2}}\big( |\Xi_{1001}\rangle - |\Xi_{0110}\rangle  \big) \ .
         \end{align}
and one quintuplet ($S=2$) state,
          \begin{align}
          |\Theta_{2,0}\rangle&=\frac{1}{\sqrt{6}}\left( |\Xi_{0011}\rangle +
          |\Xi_{0101}\rangle +
          |\Xi_{0110}\rangle + 
          |\Xi_{1001}\rangle + 
          |\Xi_{1010}\rangle +
          |\Xi_{1100}\rangle \right) \ .
         \end{align}

Instead of the computer implementation of these functions with abstract $|\alpha\rangle$ and $|\beta\rangle$ labels, we found it more convenient to use the vector representation for a general $N$-electron implementation, and especially for the computation of matrix elements with spin-dependent relativistic operators. Therefore, in this work, we have switched to the $2^N$-dimensional spinor representation. 

For the transformation between the two representations, \emph{i.e.,} the spin-adapted $\theta$-parameterisation vs. spinor linear combination, the relation between $C_{k,S}$ and $d^{(SM_S)}_{\tms_{1},...,\tms_{{\nel}}}$ is
\begin{align}
 |\chi_{S,M_S}(\boldsymbol{\theta})\rangle
 &= \sum_{k=1}^{N_\text{s}(\nel,S)}C_{k,S}(\boldsymbol{\theta})|\Theta^{(k)}_{S,M_S}\rangle \ , \nonumber  \\ 
 &= \sum_{k=1}^{N_\text{s}(\nel,S)}C_{k,S}(\boldsymbol{\theta})\sum_{\substack{\tms_1,...,\tms_{\nel} \\ \Sigma \tms_i=2M_S+1}}Q^{S,M_S,k}_{\tms_1...\tms_{\nel}} |\Xi_{\tms_1 \tms_2... \tms_{\nel}}\rangle \nonumber \\
 &=
 \sum_{\substack{\tms_{1},...,\tms_{{\nel}} \\\Sigma \tms_{i}=2M_S+1}}d^{(SM_S)}_{\tms_1,...,\tms_{\nel}}(\boldsymbol{\theta}) \, |\Xi_{\tms_1 \tms_2... \tms_{\nel}}\rangle \ , \label{eq:genexprchi}
\end{align}
and this change of representation is performed in an automated procedure in QUANTEN. For the example of the helium dimer's case, the non-relativistic energy optimisation was performed in the spin-adapted $\theta$-parameterization \cite{SuVaBook98,MaRe12}, which was converted to the spinor representation, which means storing a $2^4$-dimensional vector for every basis function (instead of 1-2 $\vartheta$ parameters of the $\theta$-parameterization, depending on the actual spin subspace). The latter representation can be more straightforwardly used to evaluate the spin-dependent Breit-Pauli (BP) matrix elements.

Although the non-relativistic wave functions are degenerate in $M_S$,  we need to generate functions with another $M_S$ value for computing the BP matrix elements. For this purpose, Eq.~\eqref{eq:genexprchi} and the matrix representation of the spin-ladder operators were employed.

\section{Higher-order fine-structure corrections from other electronic states
\label{sec:PTaA}}

In this section, we examine possible $\alpha^4\Eh$-order relativistic contributions for close-lying electronic states that can couple through the spin-dependent BP terms. At the same time, we note that a complete treatment of the $\alpha^4\Eh$ correction requires consideration of several QED terms as well \cite{DoKr74,Pa05FW,PaYe09}, which are beyond the scope of this paper.  

Closely degenerate states, such as the $\text{a}\ ^3\Sigma_\text{u}^+$ and $\text{A}\ ^1\Sigma_\text{u}^+$ states of He$_2$, can be treated in a quasi-degenerate framework. This approach takes into account the small mixing of the two states due to the perturbation. For this purpose, we build up the cumulative second-order Hamiltonian, $\mathbf{H}^{[2]}$, which includes contributions from both the direct couplings and the second-order perturbative corrections,
\begin{align}
    \mathbf{H}^{[2]} \, \mathbf{c}_i^{[2]} = E_i^{[2]} \mathbf{c}_i^{[2]} \ . 
\end{align}
The matrix element of the cumulative second-order Hamiltonian, $\mathbf{H}^{[2]}$, is given by \cite{ShBa09}:
\begin{align}
    H_{X,Y}^{[2]} = \left \langle X \left |  \hat{H}^{(0)} + \alpha^2 \hat{H}_\mathrm{sd}^{(2)} + \alpha^4 \hat{H}_\mathrm{sd}^{(2)} \frac{\hat{Q}} {E^{(0)}_Y-\hat{H}^{(0)}}\hat{H}_\mathrm{sd}^{(2)} \right| Y   \right \rangle \ , 
\end{align}
where the states 
$X$ and $Y$ are the states considered within the model space. The operator $\hat{Q}$ is the projector to orthogonal complement to the coupled ($X,Y$) subspace, \emph{e.g.,} $\hat{Q}=1- | X \rangle \langle X | - | Y \rangle \langle Y | $.

For the specific example of the 
a~$^3\Sigma_\text{u}^+$ (denoted for short by a$_{M_S}$, $M_S=0,\pm 1$)  and
A~$^1\Sigma_\text{u}^+$, the triplet states $a_{\pm 1}$ do not couple to any other states at the second-order level, so we only need to consider the diagonal elements for these states. The second-order energy correction for these triplet states is then expressed as,
\begin{align}
    E_{\mathrm{a}_{\pm1}}^{[2]}=H_{\mathrm{a}_{\pm1},\mathrm{a}_{\pm1}}^{[2]} = \left \langle \mathrm{a}_{\pm1} \left |  \hat{H}^{(0)} + \alpha^2 \hat{H}_\mathrm{sd}^{(2)} + \alpha^4 \hat{H}_\mathrm{sd}^{(2)} \frac{\hat{Q}}{E^{(0)}_\mathrm{a}-\hat{H}^{(0)}}\hat{H}_\mathrm{sd}^{(2)} \right| \mathrm{a}_{\pm1}   \right \rangle \ .
\end{align}
For A and a$_0$ there is a coupling at second order, so the matrix elements of the second-order Hamiltonian are
\begin{align}
    H_{\mathrm{a}_{0},\mathrm{a}_{0}}^{[2]} &= \left \langle \mathrm{a}_0 \left |  \hat{H}^{(0)} + \alpha^2 \hat{H}_\mathrm{sd}^{(2)} + \alpha^4 \hat{H}_\mathrm{sd}^{(2)} \frac{\hat{Q}}{E^{(0)}_\mathrm{a}-\hat{H}^{(0)}}\hat{H}_\mathrm{sd}^{(2)} \right| \mathrm{a}_0   \right \rangle \ , \\
    H_{\mathrm{A},\mathrm{A}}^{[2]} &= \left \langle \mathrm{A} \left |  \hat{H}^{(0)} + \alpha^2 \hat{H}_\mathrm{sd}^{(2)} + \alpha^4 \hat{H}_\mathrm{sd}^{(2)} \frac{\hat{Q}}{E^{(0)}_A-\hat{H}^{(0)}}\hat{H}_\mathrm{sd}^{(2)} \right| \mathrm{A}   \right \rangle \ ,
\end{align}
\begin{align}
    H_{\mathrm{a}_0,\mathrm{A}}^{[2]} &= \alpha^4 \left \langle \mathrm{a}_0 \left |  \hat{H}_\mathrm{sd}^{(2)} \frac{\hat{Q}}{E^{(0)}_\mathrm{A}-\hat{H}^{(0)}}\hat{H}_\mathrm{sd}^{(2)} \right| \mathrm{A}   \right \rangle \ , \\
    H_{\mathrm{A},\mathrm{a}_0}^{[2]} &= \alpha^4 \left \langle \mathrm{A} \left |  \hat{H}_\mathrm{sd}^{(2)} \frac{\hat{Q}}{E^{(0)}_\mathrm{a}-\hat{H}^{(0)}}\hat{H}_\mathrm{sd}^{(2)} \right| \mathrm{a}_0   \right \rangle \ .
\end{align}
The two-by-two
matrix can be diagonalized, leading to the eigenvalues,
\begin{align}
    E_{\mathrm{a}_0}^{[2]} &=    \frac{H_{\mathrm{a}_{0},\mathrm{a}_{0}}^{[2]}}{2} \left( 1+ \sqrt{1+D} \right) +  \frac{H_{\mathrm{A},\mathrm{A}}^{[2]}}{2} \left( 1- \sqrt{1+D} \right) \ , \\
    E_{\mathrm{A}}^{[2]} &=    \frac{H_{\mathrm{A},\mathrm{A}}^{[2]}}{2} \left( 1+ \sqrt{1+D} \right) +  \frac{H_{\mathrm{a}_0,\mathrm{a}_0}^{[2]}}{2} \left( 1- \sqrt{1+D} \right) \ , \\
    D &= 4 \frac{H_{\mathrm{a}_0,\mathrm{A}}^{[2]}H_{\mathrm{A},\mathrm{a}_0}^{[2]}}{\left( H_{\mathrm{a}_{0},\mathrm{a}_{0}}^{[2]} -H_{\mathrm{A},\mathrm{A}}^{[2]} \right)^2} \ .
\end{align}
Based on this formalism, we computed a numerical estimate for the second-order coupling of \atSup\ and A~$^1\Sigma_\text{u}^+$ through the f~$^3\Pi_\text{u}$  and F~$^1\Pi_\text{u}$ states, identified as the most important `interaction transmission' partners based on Ref.~\cite{EpMoChTe24}. Pilot computations suggest that the resulting second-order contribution is in the $10^{-12}\ \Eh$ range, so completely negligible in this work.

\end{document}